\documentclass[aps,pre,groupedaddress]{revtex4}

\usepackage{psfrag}
\usepackage{epsfig}
\usepackage{latexsym}

\newcommand{\bnabla}{\mbox{\boldmath $\nabla$}}
\renewcommand{\vec}[1]{\mbox{\boldmath $#1$}}

\newcommand{\pd}[1]{\partial_{#1}}
\renewcommand{\vec}[1]{\mbox{\boldmath $#1$}}

\begin{document}

\title{Drag reduction in pipe flow by optimal forcing}

\author{Ashley P. Willis}
\email{willis@ladhyx.polytechnique.fr}
\author{Yongyun Hwang}
\email{yongyun@ladhyx.polytechnique.fr}
\author{Carlo Cossu}
\email{carlo@ladhyx.polytechnique.fr}

\affiliation{
   Laboratoire d'Hydrodynamique (LadHyX),  \'Ecole Polytechnique, 91128 Palaiseau, France
}

\begin{abstract}
\noindent
In most settings, from international pipelines to home water supplies,
the drag caused by turbulence raises pumping costs
many times higher than if the flow were laminar.
Drag reduction has therefore long been an aim of high priority.
In order to achieve this end, any drag reduction method must modify
the turbulent mean flow.  Motivated by minimization of the input 
energy this requires, linearly optimal forcing functions are examined.
It is shown that the forcing mode leading to the greatest response of the flow
is always of $m=1$ azimuthal symmetry.  Little evidence is seen of the second peak at 
large $m$ (wall modes) found in analogous optimal growth calculations,
which may have implications for control strategies.
The model's prediction of large response of the large length-scale modes
is verified in full direct numerical simulation of turbulence
($Re=5300$, $Re_\tau\approx 180$).
Further, drag reduction of over 12\% is found for finite amplitude forcing 
of the largest scale mode, $m=1$.  Significantly, the forcing energy required 
is very small, being less than 2\% of that by the through pressure, resulting 
in a net energy saving of over 10\%.
\end{abstract}


\maketitle

Eddies in turbulent flow dramatically elevate the effective viscosity of
the fluid, and thereby the friction drag at the wall.
The ever increasing pressure for energy efficiency has
heightened interest in turbulent drag reduction.
Whilst strategies for drag reduction have been developed in principle,
unfortunately few have yet been realised with a net energy saving.

In wall-bounded turbulent shear flows two distinct spatial scales
become relevant at large Reynolds number: the outer scale $h$ (e.g the radius of
the pipe, or the half-width of the channel) and the inner-scale $\nu/u_\tau$,
which depends on the fluid kinematic viscosity $\nu$ and on the `wall'-velocity, 
$u_\tau=(\tau_w/\rho)^{1/2}$, based on the wall shear-stress $\tau_w$.
One of many potential difficulties is that most of the proposed strategies 
are based on the control of near-wall structures, which both change with the 
Reynolds number and are extremely small at large Reynolds numbers.
Turbulent skin friction is known to be closely associated with near-wall 
structures which scale on this inner scale \cite{choi94}.  
Understandably, there have been many investigations to manipulate these near-wall vortical structures, for example, opposition controls \cite{choi94} and spanwise oscillation of entire wall \cite{jung92,quadrio04} for plane channel flow, and \cite{choi97,quadrio00} for pipe flow.
These methods offer $\approx$7\% net energy savings, although a large part of the energy 
budget may be required to modify the flow.
%
%
Passive methods, requiring no external energetic input other than the 
driving pressure, are another promising route.  
Streamwise aligned riblets have been used to
produce energy savings, also around 7\%, much more cheaply than
via active units \cite{walsh90,bechert89}.  Unfortunately, each riblet 
must scale on the inner scale, and may require cleaning should 
the deposition of wax or dirt occur.
%


As an alternative route to the manipulation of individual near-wall structures,
it has been suggested that many could be manipulated at once 
by means of large-scale control.
The concept was proven in a nice study of the channel flow \cite{schoppa98}
by imposing large streamwise vortices $\lambda^+\approx 400$.
The spacing of near-wall structures is typically $\approx 100$ wall-units \cite{Kline67}.  
Drag reduction reaching $20$\% was achieved, and while the 
effective work done to impose the vortices was not documented, their amplitude
was only 6\% of the centre-line speed.
The feasibility of this strategy has been confirmed by an experimental
investigation \cite{iuso02} in the flat plate boundary layer where large scale
vortices were forced using vortex generators mounted on the wall.  It was 
preliminarily shown that forcing at large spanwise scale ($\lambda^+\approx 1000$) 
is also effective in reducing the drag.

It is convenient that large-scale vortices can induce drag reduction, as it is known
they can be efficiently forced efficiently in laminar flows 
\cite{butler92,Schmid2001}.
It has also been realised,
via the study of optimal perturbations,
that the emergence of large scale motions 
is important in turbulent flows
\cite{delAlamo2006,Pujals2009,Cossu2009,Hwang2009}. 
Optimal perturbations consist of streamwise uniform vortices that optimally extract energy from the mean flow while inducing the growth of the streaks. 
As the large-scale optimal vortices scale on the outer scale, they are apt to
implementation for control, as their shape is much more dependent 
on the domain geometry than on the Reynolds number.
Further, the energy input necessary to obtain a desired amplitude of streak decreases when the Reynolds number is increased.  In other words, the efficiency
of the forcing of the large scale streaks {\em increases} with the Reynolds number.

%

The scope of the present investigation is twofold.  First we compute the optimal coherent
streamwise vortices and streaks for turbulent pipe flow, in particular to determine the
associated optimal energy amplification.  In doing so, large growth is demonstrated, and
the optimal azimuthal and streamwise wavenumbers are identified.
Second, we ascertain that forcing at finite amplitude 
can induce competitive drag reduction in the turbulent pipe.
In our simulations the circumference corresponds to 
$\lambda^+_{m=1}\approx 1130$, and the forced rolls 
are significatly larger than the typical spacing of near-wall structures.
As the structure of the forced large-scale motion does not depend
strongly on the flow rate, there is promise that
a method of forcing to produce drag reduction
should be practical over large ranges of Reynolds numbers.

%

%
%
%
%

Consider the turbulent incompressible flow of a viscous fluid of kinematic viscosity $\nu$
in a circular pipe of radius $R$. The bulk velocity $U_b$ is assumed constant. Velocities are
normalised by $2U_b$ and lengths by $R$.  The Reynolds number is $Re=2\, U_bR/\nu$.
Following the approach used in \cite{delAlamo2006,Pujals2009,Cossu2009,Hwang2009},
in order to compute optimal coherent perturbations, we consider small perturbations $p$ and 
$\vec{u}$, to the pressure and turbulent mean flow $\vec{U}=U(y)\vec{\hat{z}}$, where $y=1-r$.
The perturbations satisfy continuity and the linearised momentum
equation:
\begin{eqnarray}
   \label{eq:LinearizedTurbNS}
   \pd{t}\vec{u} + 
   u_r\, \pd{r}U\vec{\hat{z}} + U \, \pd{z}\vec{u} = 
   - \bnabla p + \frac{1}{Re} \,
   \bnabla \cdot
   \left[ \nu_T(y) \left( \nabla \vec{u} +\nabla \vec{u}^T \right) \right] + \vec{f},
\end{eqnarray}
where the total total normalised effective viscosity is $\nu_T(y)=1+E(y)$ with
the $E(y)$ being the eddy viscosity. 
We use for $E(y)$ an expression originally suggested for pipe flow
by Cess \cite{cess58}, later used for channel flows by Reynolds and Tiederman \cite{reynolds67}:
\begin{eqnarray} 
   \label{eq:CessEy}
   E(y) & = & \frac{1}{2}
   {\Big\{ }
      1 + \frac{\kappa^2 \hat{R}^2 \hat{B}}{9}
      [2y-y^2]^2
      (3-4y+2y^2)^2
     \times
      \left[
         1 - \exp
         \left(
            \frac{-y \hat{R} \sqrt{\hat{B}}}{A^+}
         \right)
      \right]
   {\Big\} }^{\frac{1}{2}} - \frac{1}{2} \, .
\end{eqnarray} 
with  $\hat{R}=Re\,/\,2$,\, $\hat{B}=2\,B$ and the fitting parameters $A^+=27$ and $\kappa=0.42$
chosen to better match the observations, more recently studied in \cite{mckeon05}.
The mean streamwise velocity $U(y)$ in equilibrium with this eddy viscosity is easily retrieved 
from the mean averaged momentum equation equation
\begin{equation}
   \label{eq:avgedNS}
   0 = -\pd{z}P + \frac{1}{Re}
   \left(
      \frac{1}{r} + \pd{r}
   \right)
   (\nu_T \pd{r} U ),
   \qquad
   B = -\pd{z}P .
\end{equation}

The symmetry of the base flow allows one to consider separately 
perturbations of the form 
$\vec{u}(r,\theta,z;t)=\hat{\vec{u}}(r,m,\alpha;t)
\,{\rm e}^{{\rm i}(\alpha z + m \theta)}$, where
$\alpha$ and $m$ are the streamwise and azimuthal wavenumbers respectively.
The eigenvalues and eigenvectors are found directly
from the linearised
primitive variable system with explicit solenoidal condition
\footnote{
 Calculations were performed using a Chebyshev collocation method
 with $N$ points on $r\in[0,1]$ .
 The primitive variable eigensystem avoids high order derivatives
 but has infinite eigenvalues.  The eigensolver, however,
 returns a predictable number of these which are easily filtered.
 To be safe, 95\% of the ($2N-3$) eigenfunctions returned
 were kept for the analysis of optimals.
 At the highest $Re=10^6$ with $N=250$
 the power spectral drop-off of the optimal mode was of $8$ orders of magnitude
}.
For all calculations considered $U(y)$ is found to be linearly stable, in accordance with similar
results obtained in the plane channel and boundary layer flows \cite{reynolds67,delAlamo2006,Cossu2009,Pujals2009,Hwang2009}.

\begin{figure}
      \psfrag{m=1}{\tiny $m=1$}
      \psfrag{a=0}{\tiny $\alpha=0$}
      \psfrag{a=1}{\tiny $\alpha=1$}
      \psfrag{Rmax}{\small $R_{\mathrm{max}}$}
      \psfrag{m}{\small $m$}
      \psfrag{R(w)}{\small $R(\Omega_f)$}
      \psfrag{w}{\small $\Omega_f$}
      \psfrag{Re}{\small $Re$}
      \epsfig{figure=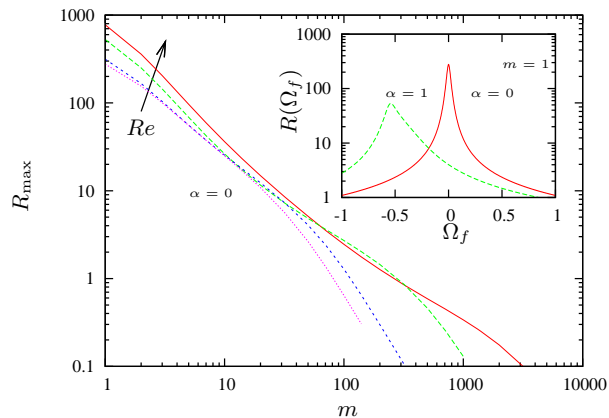, angle=0, scale=0.95}
   \caption{\label{fig:Rmax}
      Response $R_{\mathrm{max}}(\alpha=0,m)$ to forcing,
      $Re=5300$, $10^4$, $10^5$, $10^6$, the arrow indicating increasing $Re$.
      Inset: $Re=5300$, $m=1$, response to forcing of frequency $\Omega_f$.
      The turbulent flow is most responsive to axially-independent steady
      forcing on the largest scale.
   }
\end{figure}

The optimal response to forcing is calculated using methods
described in \cite{schmid07}.
For a harmonic
forcing of the form
$\vec{\hat{f}}(r,m,\alpha;t) 
= \vec{\tilde{f}}(r,m,\alpha)\, {\mathrm e}^{{\mathrm i}\,\Omega_f t}$ 
and response
$\vec{\hat{u}}(r,m,\alpha;t) 
= \vec{\tilde{u}}(r,m,\alpha)\, {\mathrm e}^{{\mathrm i}\,\Omega_f t}$ 
the optimal response is given by
\begin{equation}
   R(\alpha,m;\Omega_f)  = \max_{\vec{\tilde{f}}} \,
   \frac{||\vec{\tilde{u}}||}{||\vec{\tilde{f}}||} \, ,
\quad
   R_{\mathrm{max}}(\alpha,m)  = \max_{\Omega_f} R(\Omega_f) .
\end{equation}
Figure \ref{fig:Rmax} shows our calculations for the optimal response to forcing
for several $Re$
\footnote{
   For $Re_\tau=u_\tau R/\nu$ where
   $u_\tau=(\nu\,\pd{r}U(r))^{\frac{1}{2}}|_{r=R}$,
   $Re=5300,10^4,10^5,10^6$ $\to$ $Re_\tau=180,317,2380,19200$.
}.  
The optimal response $R_{\mathrm{max}}$ generally increases with increasing $Re$.
As the spanwise length-scale is decreased ($m$ increased) the
response rapidly decreases.  The largest possible response occurs
for the mode $m=1$.
Huge response to forcing of large scale modes is possible.
The relative difficulty of forcing motion on small scales, on the other hand,
implies that considerable effort is required to locally
manipulate structures in the neighbourhood of the wall.

The analysis above is linear and uses the eddy viscosity assumption.
A first step towards application is to verify that large responses
are obtained in fully nonlinear turbulent numerical simulation.
The pipe flow code described in \cite{willis09} has been used for
full simulation of turbulence subject to the optimal forcing
$\vec{\tilde{f}}$.
The numerical method uses a Fourier decomposition in $\theta$ and
$z$ and finite differences in $r$
\footnote{
   Minimum and maximum radial spacings
   of points are $0.11$ and $4.4$ wall units, $\nu/u_\tau$.
   Points are separated by $5.9$ in $\theta$ and $9.4$ in $z$ at $r=R$ and $L=15\,R$.
   These values are $0.08$, $4.0$, $3.9$, $6.3$, $20$
   respectively for the larger calculation.
}.
For the following calculations
the computational domain is of length
$L=15R$, chosen to include the very large-scale motions reported
in \cite{kim99} for $L$ up to $14$.
We choose to perform the simulations enforcing a fixed flux,
consistent with previous analyses \cite{schoppa98,willis09}.

\begin{figure}
   \begin{center}
      \psfrag{t}{\small $t$}
      \psfrag{Cf/Cf0}{\small $C_f/C_f^0$}
      \epsfig{figure=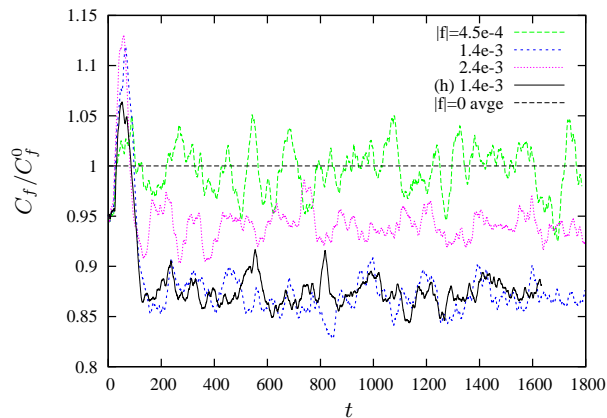, angle=0, scale=0.95}
   \end{center}
   \caption{\label{fig:5300Rspnc}
      Relative drag for several $||\vec{\tilde{f}}||$ at $Re=5300$.
      The straight horizontal line is the unforced average. 
      The solid line is a higher resolution check for the best case.
   }
\end{figure}
Figure \ref{fig:5300Rspnc} shows time series for the skin-friction coefficient,
$C_f=2(u_\tau/U_b)^2$, for three levels of forcing of the $m=1$ mode at $Re=5300$.
The horizontal line is the time average $C_f$ for zero forcing.
The solid line for a higher resolution calculation verifies no
loss of the observed drag reduction for the case shown.

The response to forcing in simulations is measured by 
$R'=||\vec{\bar{u}}||\,/\,||\vec{\tilde{f}}||$,
where the averaged response field is
$\bar{\vec{u}}(r,\theta) = \langle\vec{u}\rangle_{z,t} - U(r)\,\hat{\vec{z}}$ 
and $U(r)= \langle\vec{u}\rangle_{\theta,z,t}$ is the mean flow in the absence of forcing.
The norm used here is 
$||\vec{a}||^2=\frac{1}{L}\int \vec{a}\cdot\vec{a}\,\mathrm{d}V$ and 
$\langle\,\cdot\,\rangle_s$ indicates averaging over the subscripted variables.
Time averages are taken over $1400\,R/(2U_b)$ time units.
Measures of streak and roll components of the flow are
$A_z = \frac{1}{2}(\max \bar{u}_z - \min \bar{u}_z)$ and
$A_h = \max |(\bar{u}_r,\bar{u}_\theta,0)|$.

In table \ref{tab:runs} it is seen that the response to forcing may indeed be large,
in agreement with the model.  The prediction from the linear model is $R_{\mathrm{max}}=276$
for $m=1$, and there remains a factor $2$ difference, however, with the simulations for 
the two smallest $||\vec{\tilde{f}}||$.  As the time-averaged mean flow for the unforced case,
obtained from numerical simulation, 
is close to that given by (\ref{eq:CessEy}) and (\ref{eq:avgedNS}),
the difference must originate from the isotropic eddy viscosity assumption.
While at first both $A_z$ and $A_h$ increase linearly with
$||\vec{\tilde{f}}||$, streaks saturate near 
19\% of the centre-line speed.  Streak saturation occurs when the roll amplitude 
is approximately 7\% of the centre-line speed, 
and coincides with when the maximum drag reduction is achieved.
\begin{table}
  \begin{tabular}{|crccrrcc|}
   \hline
   $||\vec{f}||$ & $R'$ &  $P$ & $W$ &
   $\mathrm{DR}$ & $E_\mathrm{net}$ & $A_h/U_{cl}$ & $A_z/U_{cl}$ \\
   \hline
   ($m=1$) & & & & & & & \\
      1.4e-4 & 146.0 &  3.680e-3 & 1.03e-6 &  0.2 &  0.2 & 0.0091 & 0.0372 \\
      4.5e-4 & 138.0 &  3.675e-3 & 1.10e-5 &  0.4 &  0.1 & 0.0217 & 0.1088 \\
      7.7e-4 & 109.7 &  3.456e-3 & 3.29e-5 &  6.3 &  5.4 & 0.0444 & 0.1620 \\
      1.0e-3 &  90.3 &  3.230e-3 & 5.02e-5 & 12.4 & {\bf 11.1} & 0.0540 & 0.1847 \\
      1.2e-3 &  77.1 &  3.222e-3 & 6.53e-5 & 12.6 & 10.9 & 0.0618 & 0.1884 \\
      1.4e-3 &  63.3 &  3.218e-3 & 8.77e-5 & {\bf 12.8} & 10.4 & 0.0740 & 0.1890 \\
      2.0e-3 &  41.4 &  3.356e-3 & 1.57e-4 &  9.0 &  4.8 & 0.1049 & 0.1899 \\
      2.4e-3 &  33.1 &  3.470e-3 & 2.24e-4 &  5.9 & -0.1 & 0.1308 & 0.1946 \\
      4.5e-3 &  16.3 &  3.990e-3 & 5.58e-4 & -8.1 &-23.3 & 0.1886 & 0.1953 \\[2pt]
   ($m=2$) & & & & & & & \\
      7.7e-4 & 118.0 &  3.358e-3 & 2.40e-5 &  9.0 &  8.3 & 0.0355 & 0.1605 \\
      1.4e-3 & 102.4 &  3.306e-3 & 7.81e-5 & 10.4 &  8.3 & 0.0699 & 0.2608 \\
      2.4e-3 &  62.7 &  3.660e-3 & 1.75e-4 &  0.8 & -4.0 & 0.0910 & 0.2691 \\[2pt]
   ($m=4$) & & & & & & & \\
      4.5e-4 & 130.6 &  3.557e-3 & 9.66e-6 &  3.6 &  3.3 & 0.0224 & 0.1133 \\
      7.7e-4 & 137.9 &  3.540e-3 & 2.87e-5 &  4.0 &  3.2 & 0.0461 & 0.2208 \\
      1.4e-3 &  92.0 &  3.958e-3 & 6.79e-5 & -7.3 & -9.1 & 0.0653 & 0.2638 \\[2pt]
   ($m=12$) & & & & & & & \\
      7.7e-5 &  46.4 &  3.702e-3 & 8.64e-8 & -0.3 & -0.3 & 0.0012 & 0.0072 \\
      2.4e-4 &  46.3 &  3.714e-3 & 7.07e-7 & -0.7 & -0.7 & 0.0034 & 0.0183 \\
      7.7e-4 &  24.6 &  3.951e-3 & 6.35e-6 & -7.1 & -7.3 & 0.0099 & 0.0494 \\ 
  \hline
  \end{tabular}
  \caption{\label{tab:runs}
     Time averaged statistics for turbulent pipe flow subject to
     forcing.  $P^0=$3.689e-3, $U_{cl}=U(r=0)=0.6556$.
     Largest drag reduction and net energy reduction in bold. 
  }
\end{table}

\begin{figure}
      \psfrag{Cf/Cf0}{\small $C_f/C_f^0$}
      \psfrag{|f|}{\small $||\vec{\tilde{f}}||$}
      \epsfig{figure=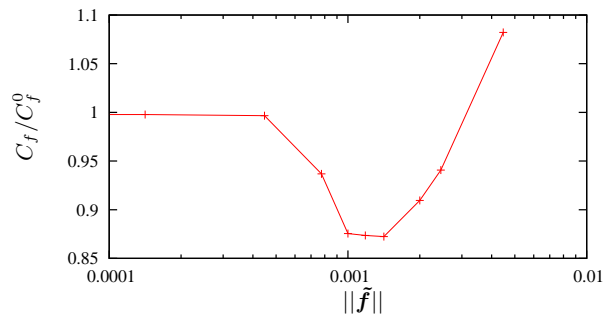, angle=0, scale=0.95}
   \caption{\label{fig:Cf_f}
      Effect of large-scale forcing, $m=1$, on the skin friction, $C_f$.
   }
\end{figure}

Also in table \ref{tab:runs} is the percentage drag reduction
$\mathrm{DR}=(100/C_f^0)(C_f^0-C_f)$,
where superscripts $0$ indicate the unforced case.
The power to drive the flow per unit length is
$P = \frac{\pi}{8} C_f$
in units $\rho\,(2U_b)^3\,R$.
The relative change in $C_f$ for increased forcing is seen in figure
\ref{fig:Cf_f}.
The drag reduction is up to $12.8$\%
and the greatest net energy saving up to $11.1$\%, where
$E_{\mathrm{net}}=(100/P^0)[P^0-(P+W)]$.
For the greatest $E_{\mathrm{net}}$, the power consumption by
forcing,
$W = \frac{1}{L}\int \vec{u}\cdot\vec{\tilde{f}}\,\mathrm{d}V$,
is only $1.4$\% of the
power required to maintain the flux of
the unforced flow, $P^0$, or $1.6$\% of that to
drive the resulting flow.

\begin{figure}
   \begin{tabular}{ccc}
      \epsfig{figure=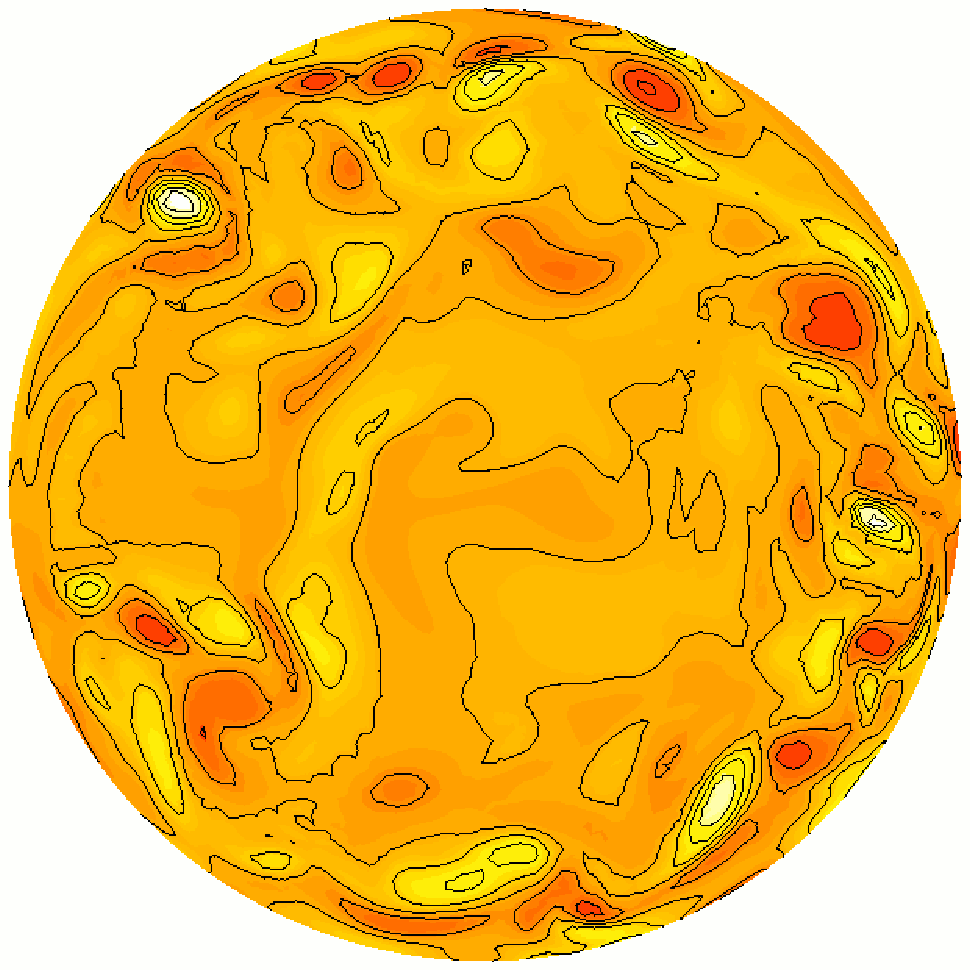,    angle=0, width=26mm} &
      \epsfig{figure=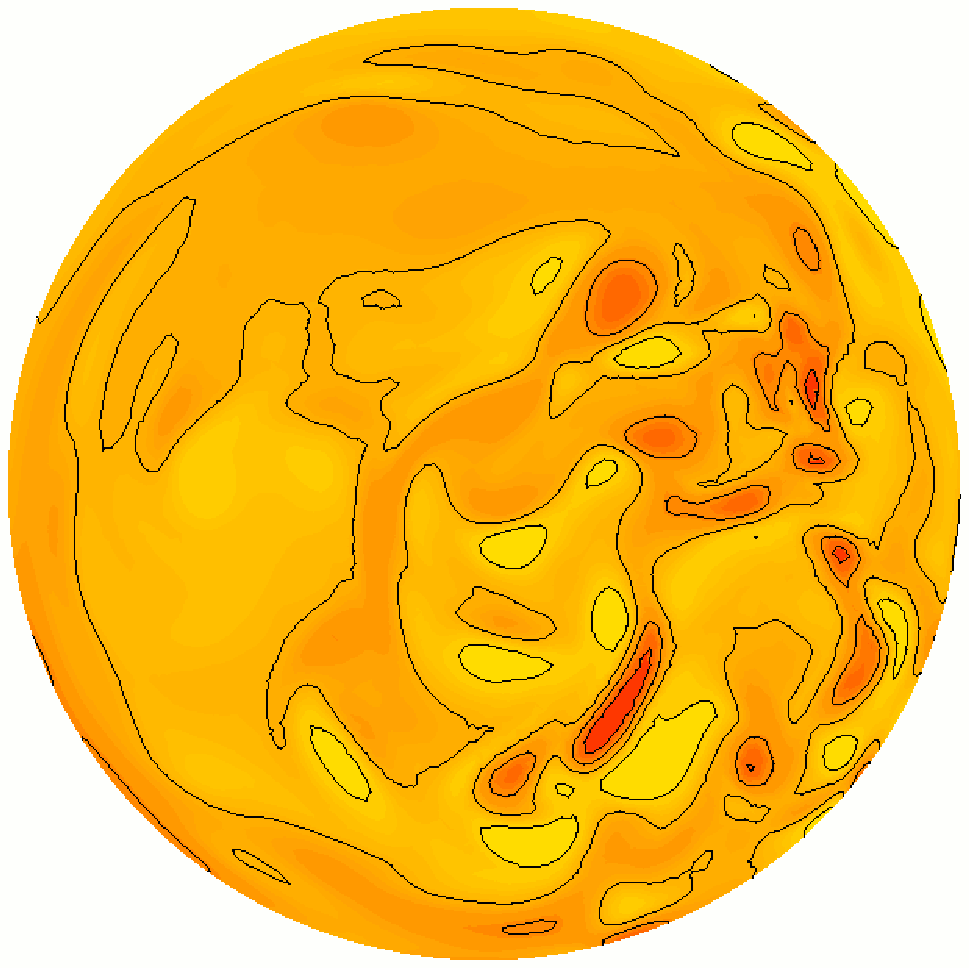, angle=0, width=26mm} &
      \epsfig{figure=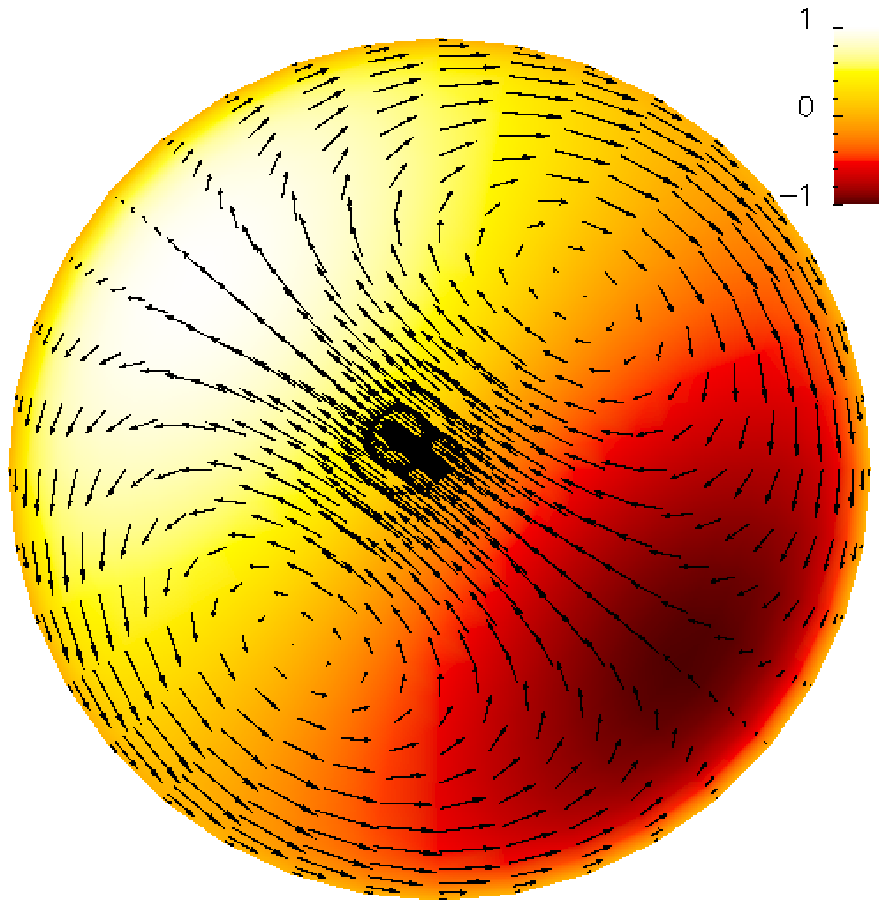,    angle=0, width=27mm} \\
      {\small (a)} & 
      {\small (b)} & 
      {\small (c)} \\
      \epsfig{figure=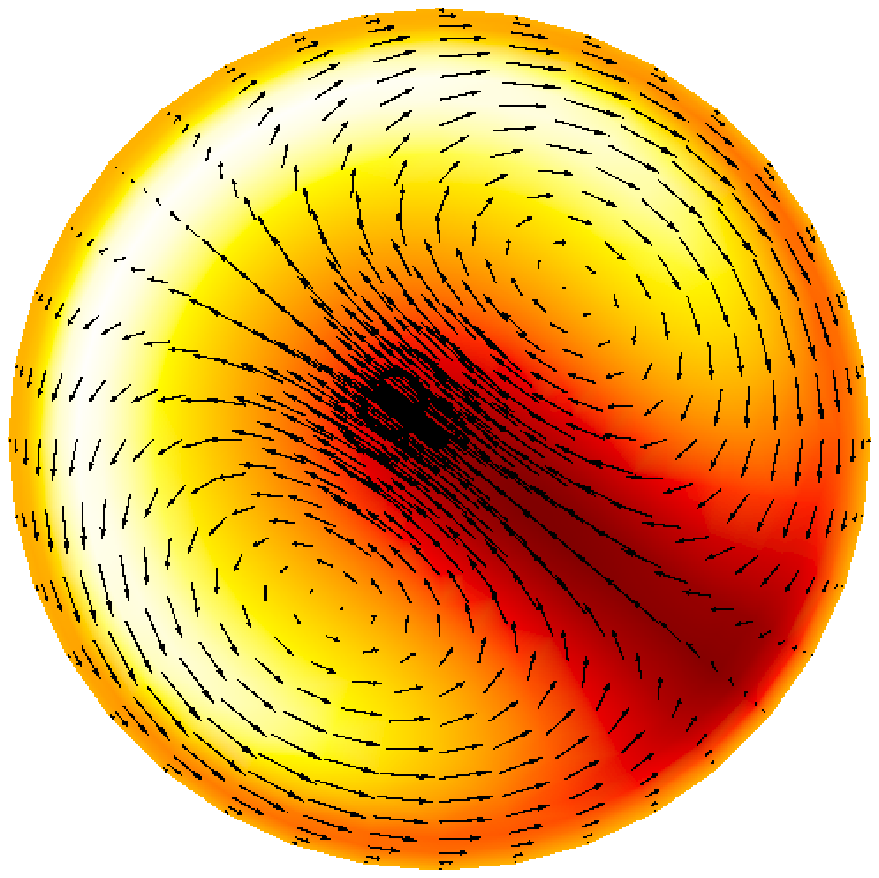,       angle=0, width=27mm} &
      \epsfig{figure=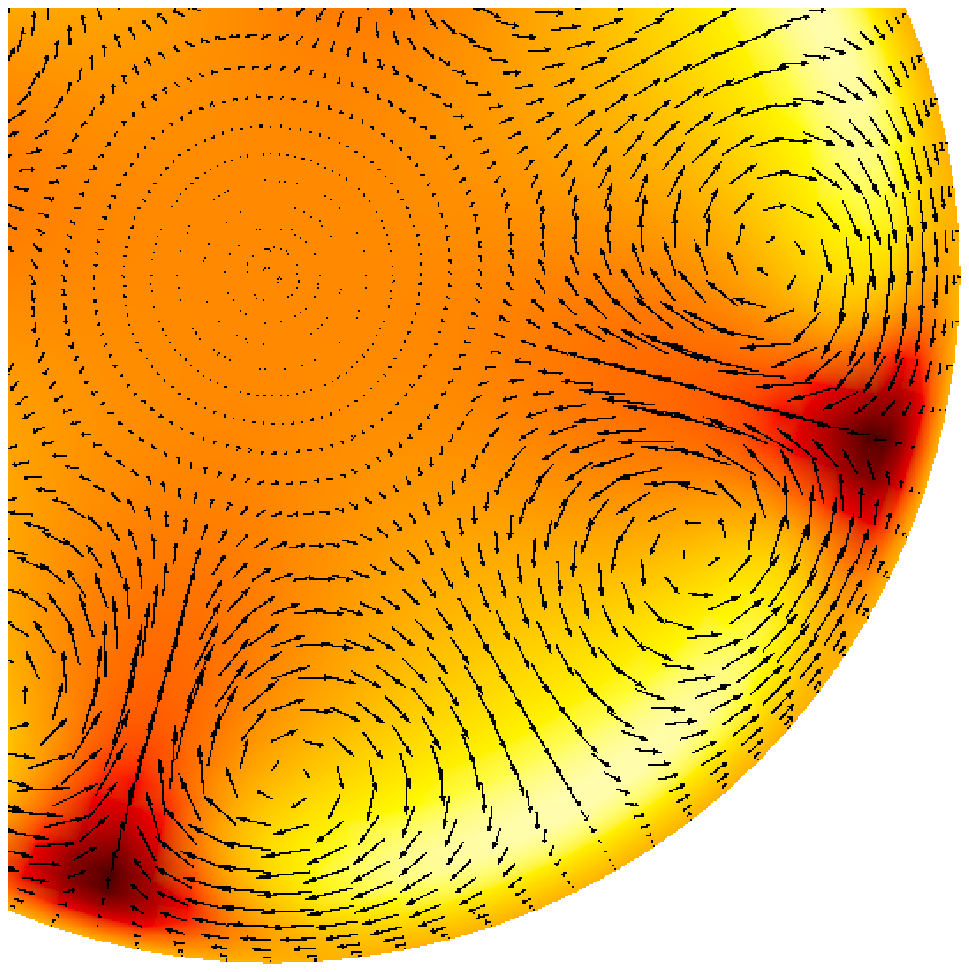,      angle=0, width=24mm} &
      \epsfig{figure=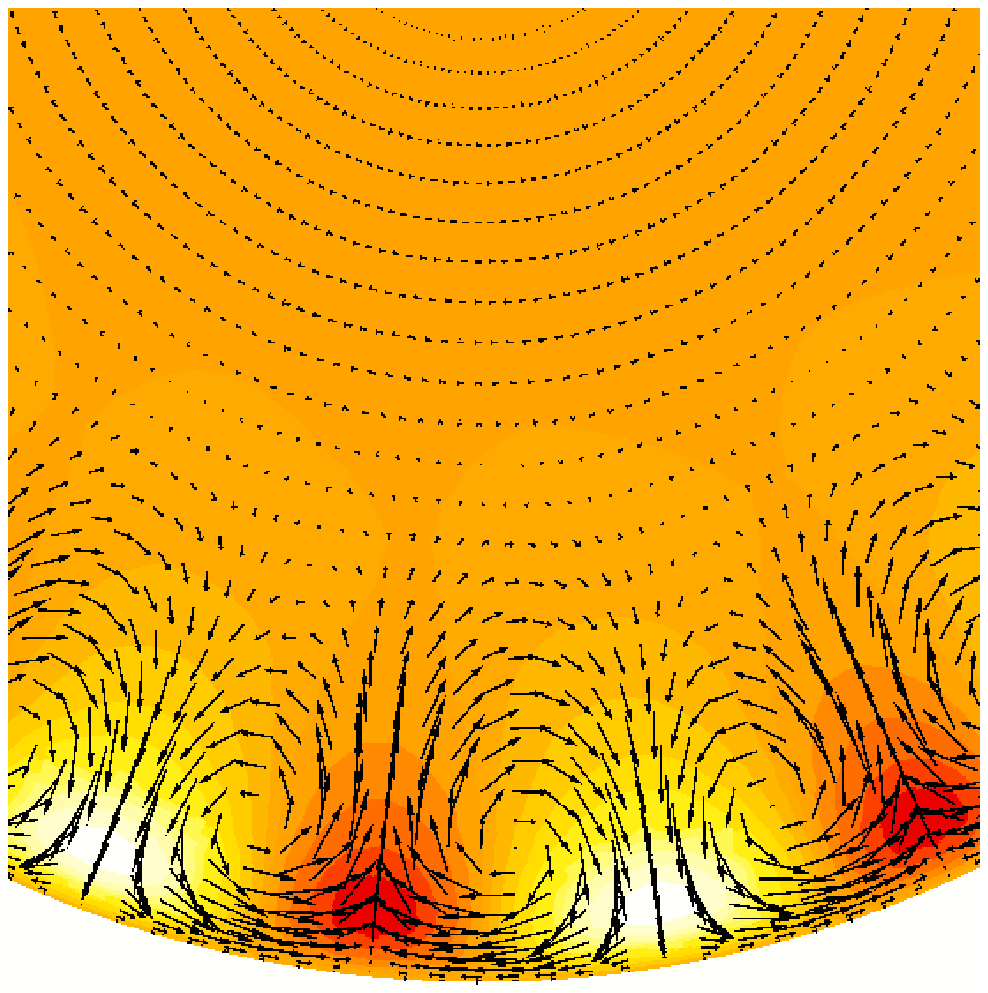,      angle=0, width=24mm} \\
      {\small (d)} &
      {\small (e)} & 
      {\small (f)} 
   \end{tabular}
   \caption{\label{fig:fModRes}
      (a) Snapshot of axial vorticity with no forcing;
      (b) Snapshot on same scale as (a) for $||f||=$1.4e-3.
      (c) Arrows of the optimal forcing and contours of the optimal response;
      (d)-(f) Time averaged velocity for $||f||=$1.4e-3, 7.7e-4, 2.4e-4 for
      $m=1,4,12$ respectively, relative to $U(r)$ without forcing.
   }
\end{figure}
The effect of forcing on the flow is shown in figure \ref{fig:fModRes}.
Small scale axial vortices are weakened throughout the domain, and are almost suppressed in
one half.  Being largely responsible for the turbulent skin friction \cite{kravchenko93},
this indicates that near-wall turbulent production is suppressed by 
forcing large-scale outer motions, consistent with \cite{schoppa98}.
The large scale forcing structure is reproduced well in the roll 
structure of the time averaged flow.  The broad slow streak, however,
is clearly much narrower in the nonlinear response.
This narrower slow streak means that several may be packed into the 
domain (e.g.\ $m=2,4$) before
they affect each other ($m=12$).  
The case $m=12$ corresponds to the natural mean spacing
of near wall structures $\lambda^+\approx 100$.  For this case the response $R'$ 
is significantly reduced, and drag only increases as the forcing is increased.




In summary, the model successfully predicts the
large response to forcing 
observed in full simulation of turbulence.
The largest response to forcing occurs
for large-scale modes, which may have implications 
for control.
Drag reduction of over 12\%
has been shown in full simulations when
forcing the large-scale mode.  Significantly,
this was shown to be possible using very small input 
energy, being less than 2\% of the energy used
to drive the flow.

Several of the alternative methods described above predict 
much larger drag reduction, but at much greater energetic 
expense.  This is possibly linked to the small response
to forcing of near-wall vortices.
The net energy saving is similar to that here, but relies 
on both the large drag reduction being realised at the same 
time as efficient actuation.
For useful implementation of the method proposed in this study,
the induction of rolls need not be efficient. For example,
efficiency of 50\%, still using only
4\% of the driving energy, might be acceptable.
Note also that the response is expected to increase with
increasing $Re$, thereby reducing this cost.
In practise, induction of large scale rolls is
possible via passive actuators \cite{Fransson2004,Fransson2006} 
or active jets \cite{iuso02}.  Although
neither would reproduce the precise details of forcing used
here, linear optimals are known to be robust to perturbations of the
system, and responses close to those seen here can be expected.
Slightly different responses, however, may also be beneficial ---
whilst the forcing in this numerical experiment
is optimal for modifying the flow, it was not chosen to be
optimal for drag reduction in the nonlinear r\'egime.
Thus there is scope for further enhancement.


%
\bibliography{trans,carlo}
\end{document}